\documentclass{aa}
\usepackage{natbib}
\newcommand{\bepo}{{\em BeppoSAX\ }}
\newcommand{\rxte}{{\em RXTE\ }}

\newcommand{\integral}{{\em INTEGRAL\ }}
\newcommand{\xmm}{{\em XMM-Newton\ }}

\usepackage{graphicx}
\usepackage{amsmath}
\usepackage{amssymb}

\title{Long-term pulse profile study of the Be/X-ray pulsar \\ SAX J2103.5+4545}
\author{A. Camero Arranz
     \inst{1}
     \and
     C.A. Wilson
     \inst{2}
     \and
     M.H. Finger     
      \inst{3}
     \and
     V. Reglero
     \inst{1}
     }
     \institute{$^1$  GACE, Instituto de Ciencias de los Materiales, Universidad de
     Valencia, P.O. Box 20085, 46071 Valencia, Spain\\
$^2$ NASA Marshall Space Flight Center, Huntsville, AL 35812, USA \\
$^3$ National Space Science Technology Center, Huntsville, AL 35805, USA\\
$^3$ Universities Space Research Association, Huntsville, AL 35805, USA}  

\authorrunning{Camero Arranz et al.}
\titlerunning{Long-term pulse profile study of the Be/X-ray pulsar SAX J2103.5+4545}
\offprints{A. Camero Arranz\\  \email{Ascension.Camero@uv.es}}
\date{Received ; accepted   }

\begin{document}

\abstract{}
{We present the first long-term pulse profile study of the X-ray pulsar SAX J2103.5+4545. Our 
main goal is to study the pulse shape correlation either with luminosity, time or energy. 
}
{This Be/X-ray binary system was observed from 1999 to 2004 by  \rxte PCA, 
and by \integral  from 2002 to 2005, during the Performance and Verification 
(PV) phase and the Galactic Plane Scan survey (GPS). X-ray pulse profiles were obtained 
in different energy ranges. The long-term spectral variability of this source is studied. 
The long-term flux, frequency and spin-up rate histories are computed. A new set of orbital 
parameters are also determined.
}
{The pulse shape is complex and highly variable either with time or luminosity.  However, an 
energy dependence pattern was found. Single, double, triple or even quadruple  
peaks pulse profile structure  was obtained. It was confirmed that SAX J2103.5+4545
becomes harder when the flux is higher. The new  orbital solution  obtained is:\linebreak
P$_{orb}$= 12.66528$\pm$0.00051 days,  e = 0.401$\pm$0.018,  $\omega$ = 241.36$\pm$2.18 and 
a$_x$sin{\it i} =  80.81$\pm$0.67 lt-s.  
}
{}
{}

\maketitle

\section{INTRODUCTION}

SAX J2103.5+4545 is a Be High Mass X-ray Binary 
pulsar with an orbital period  of  12.68 d (the shortest known in Be/X-ray 
binary systems) and X-ray pulsations of 358 s. These systems  
consist of  a neutron star  orbiting a Be companion, hence forming a 
Be/X-ray binary. Be stars are  rapidly rotating objects with a 
quasi-Keplerian disk around their equator. The optical and infrared emission 
is dominated by the donor  star and its equatorial disk, and it is 
characterized by spectral lines in emission  (particularly 
those of  the Balmer series) and IR excess. The standard model of a 
Be/X-ray binary ascribes the high-energy radiation to an accreting 
mechanism that takes place when the compact object interacts with the Be 
star's circumstellar disk, giving rise to an X-ray outburst.

This source was discovered using the \bepo satellite \citep{Hulleman}. The X--ray 
spectrum (2--25 keV) was fitted by a power-law with 
a photon index of 1.27$\pm$0.14 plus a photoelectric absorption at lower energies 
(N$_H$=3.1$\times10^{22}$ cm$^{-2}$). The likely optical 
counterpart is a B0Ve star (V = 14.2) at a 
distance of 6.5 kpc \citep{Reig}. Using \xmm  data a quasi-periodic 
oscillation at 22.7 s was discovered by \.{I}nam et al. (2004). Baykal et al. (2002) 
found  a correlation between spin-up rate and X--ray flux 
during the 1999 outburst. This suggests the formation of an accretion disk during 
periastron passage of the neutron star. However,\linebreak SAX J2103.5+4545 does not  follow 
the Corbet P$_{orb}$/P$_{spin}$ correlation found in other accreting pulsar systems. 

Previously,  this source has been analyzed by  Hulleman et al. ({\it BeppoSAX}, 1998), 
Baykal et al. $\&$ \.{I}nam et al. ({\it RXTE}, {\it XMM-Newton}, 2000, 2004, 2007), Reig et al.
(optical \& IR wavelengths, 2004, 2005), Filippova et al. (optical 
wavelength; {\it RXTE}, {\it XMM-Newton}, {\it INTEGRAL}, 2004),  
 Blay et al.,  Falanga et al. and Sidoli et al. only with \integral data (2004, 2005, 2005,
respectively), Blay (optical/X-ray correlation; \integral 2006) and  Camero-Arranz et 
al. 2006 with \rxte $\&$ \integral. 

We present a long-term pulse profile study,  as well as a spectral and  pulse-timing 
analysis of  SAX J2103.5+4545 using  data  from both  \integral and \rxte missions.

\section{OBSERVATION AND DATA REDUCTION}
 
The \textbf{INTE}rnational \textbf{G}amma-\textbf{R}ay \textbf{A}strophysics  
\textbf{L}aboratory ({\it INTEGRAL}, Winkler et al. 2003) consists of three  
coded mask telescopes, the spectrometer SPI (20 keV--8 MeV), the imager IBIS 
(15 keV--10 MeV), and the X-ray monitor JEM-X (4--35\rm\,keV), as well as the 
optical monitoring camera OMC (V,  500--600nm).
 
SAX J2103.5+4545 has been detected  by IBIS/ISGRI during \integral PV phase and the GPS survey  
from 2002 to 2005 ( MJD 52618 -- MJD 53356),  with a total observing time of
$\sim$860 ks.  \integral data reduction was carried out with  the 
Offline Scientific Analysis software, release 5.0,  distributed
by the ISDC  \citep{Courvoisier}. A software description can be found in 
Goldwurm et al. (2003), Diehl et al. (2003), Wetergaard et al. (2003).

\rxte (Bradt et al. 1993) carries 3 instruments on board. The  
Proportional Counter Array (PCA; Jahoda et al. 1996) 
is sensitive from 2--60 keV. The High Energy X-ray Timing 
Experiment (HEXTE; Gruber et al. 1996) extends the X-ray sensitivity 
up to 200 keV. Monitoring the long-term behavior 
of some of the brightest X-ray sources, the All Sky Monitor 
(ASM; Levine et al.1996) scans most of the sky every 
1.5 hours at 2--10 keV. 

SAX J2103.5+4545 was also observed with  \rxte PCA and HEXTE from 1999 to 2004 
(MJD 51512.8 -- MJD 53047.9), with a total observing time 
of $\sim$1900 ks. However, it is important to note that during that period there were
two gaps with no observations taken by those instruments 
(MJD 51900 -- MJD 52000 and MJD 52025 -- MJD 52430).  For each available observation, we have analyzed PCA 
Standard1  data (0.125 s time resolution, no energy resolution) for the light
curves and Standard2 data (16 s time resolution, 129 channel energy
resolution) for spectral analysis using FTOOLs V6.0.5. In addition, GoodXenon
PCA data were selected for the pulse profile study, and  HEXTE binned mode data
from clusters A and B for the spectral analysis.

 \begin{figure}
  \includegraphics[width=9cm,height=14.5cm]{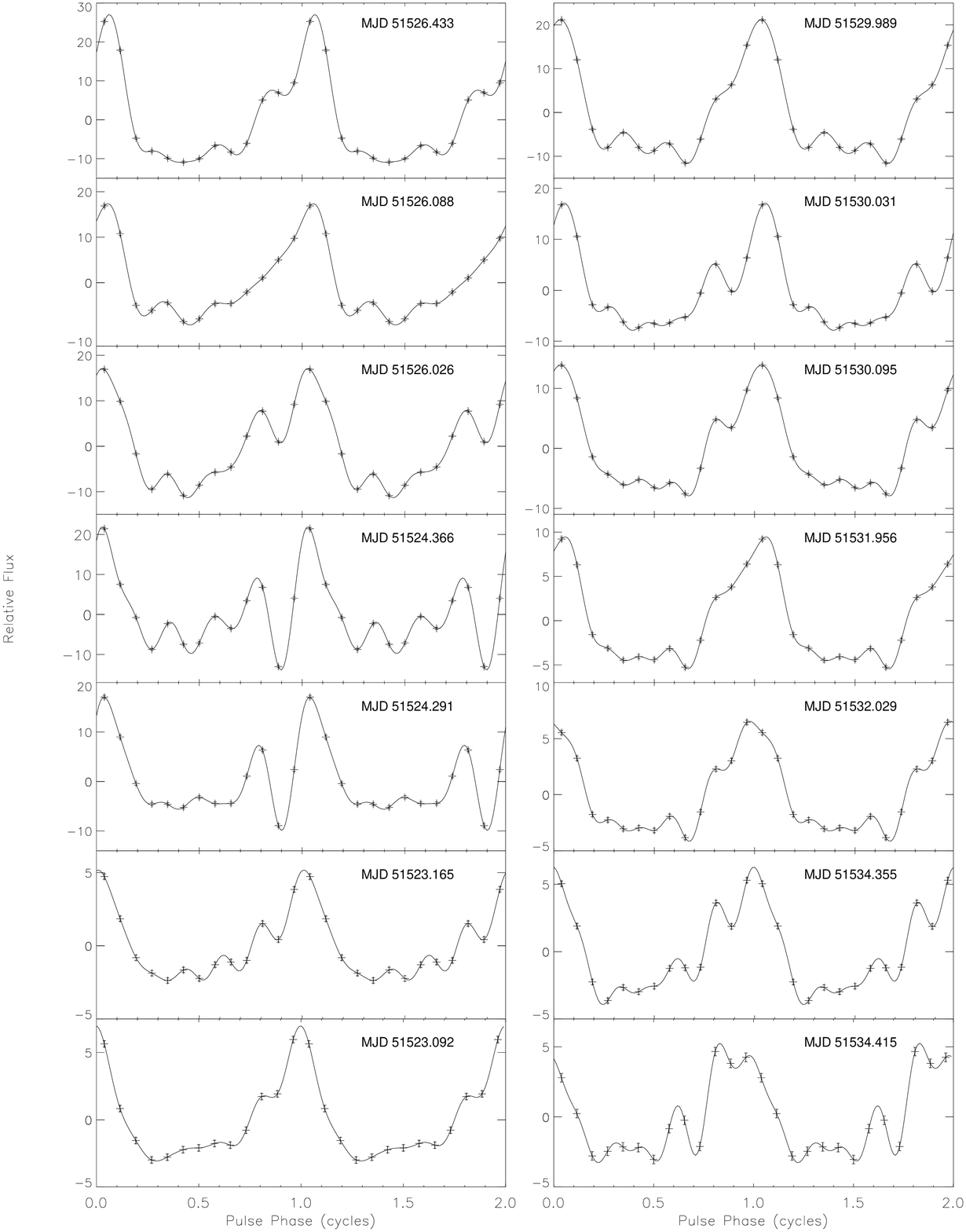}
   \caption{\rxte PCA (2--60 keV) pulse profiles consecutive in time during outburst. Two cycles are plotted
            for clarity. The evolution on intensity of the first part of the outburst is plotted on the left column
            increasing from bottom to top. The fall is plotted on the right, with intensity decreasing from 
            top to bottom. \label{consecut_prof}}
 \end{figure}

 \begin{figure}
  \includegraphics[width=8.65cm,height=4.8cm]{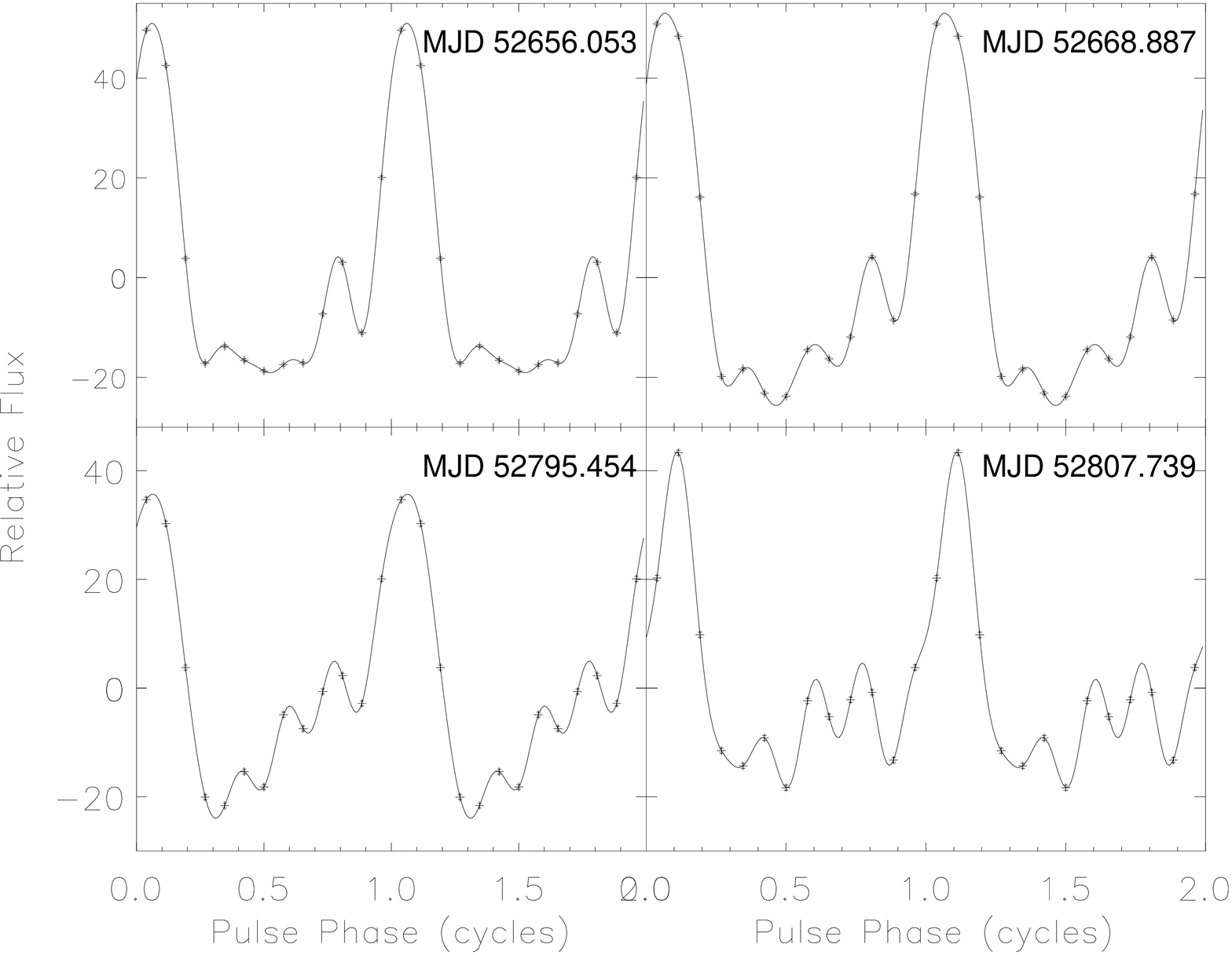}
 \caption{PCA (2--60 keV) pulse profiles at the same orbital phase (0.73$\pm$0.02 cycles)
          and at the same luminosity state  (5.62$\pm$0.04$\times$10$^{+36}$ erg  s$^{-1}$).\label{same_orb_pha_prof}}
  \end{figure}

 \begin{figure*}[!t]
  \includegraphics[width=18.8cm,height=3cm]{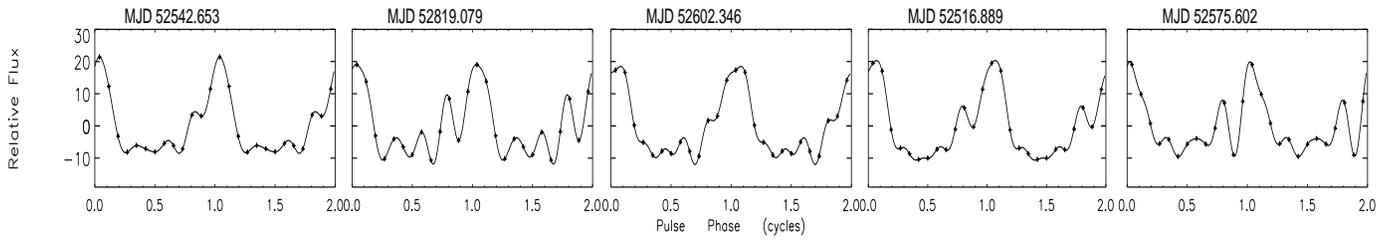}
   \caption{\rxte PCA 2--60 keV pulse profiles at the same luminosity state (L = 2$\pm$0.6$\times$10$^{+36}$ erg  s$^{-1}$).
    \label{same_lumin_prof}}
 \end{figure*}

 \begin{figure}
  \includegraphics[width=8.6cm,height=8.3cm]{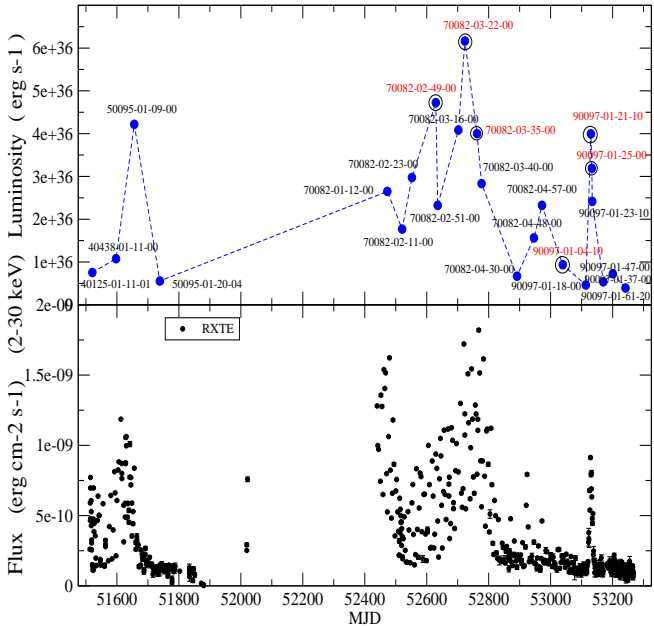}
    \caption{Top. Sample of  PCA observations selected for the pulse profile intensity analysis. 
     An extra circle indicates the existence in addition of
     an \integral IBIS/ISGRI observation. Bottom. \rxte PCA flux in the 2-30 keV energy range. \label{Lumin_vs_time}}
 \end{figure}

\begin{table} 
\centering  
  \caption{Pulse Period detections of SAX J2103.5+4545 with IBIS/ISGRI. The second column shows
        the number of ScWs used for the computation in each epoch. } 
  \begin{tabular}{lll} 
   \noalign{\smallskip}
    \hline \hline \noalign{\smallskip}
     \noalign{\smallskip}
      Epoch    &  number  &  Pulse Period  \\
     (MJD)    &  of ScWs  &  (s) \\
   \noalign{\smallskip}
   \hline \hline \noalign{\smallskip}
  52619.299  &   6  &        --           \\                  
   52630.755  &  49	&  354.927 $\pm$0.014 \\        
    52637.303  &  18  &        --      \\
    52722.884  &   3	&  354.072$\pm$0.014 \\ 
   52761.947  &  35  &  353.478$\pm$0.017\\
    52797.040  &   2  &         --  \\
   52805.978  &   3  &         --  \\
    52820.945  &   2  &         --  \\
    53021.331  &  74  &         --   \\
   53038.548  &  119 &   352.446$\pm$0.019 \\ 
    53041.996  &   21 &         --    \\
    53102.514  &   3  &         --    \\
    53130.288  &   77 &   352.350$\pm$0.006  \\ 
    53133.176  &   51 &      352.29$\pm$0.01  \\
   53188.993  &   2  &         --           \\ 
   53258.517  &   3  &         --            \\
    53333.056  &   6  &         --            \\
   53349.280  &   5  &         --            \\
    53354.965  &   4  &         --             \\
 \noalign{\smallskip}
\hline \hline \noalign{\smallskip} 
   \end{tabular} 
   \label{broad} 
\end{table} 

\section{DATA ANALYSIS AND RESULTS}
 \subsection{PULSE PROFILE}

In order to study the long-term pulse profile dependence on luminosity, time or orbital
phase we extracted 548 \rxte PCA Standard1 light curves 
in the 2--60 keV energy range. The background was not subtracted in this first 
step.  Then, we corrected the times to the barycenter of the solar 
system,  as well as for the orbital motion using the binary orbital parameters by 
Baykal et al. (2002). We constructed  pulse profiles by fitting the data with a 
harmonic expansion in pulse phase (6 harmonics were used). Typically we
use data spanning a 4000 s interval ( $\sim$11 pulse periods). Initially we used a  
simple phase model ($\phi$(t$_k$)= $\phi$$_o$+f$_o$(t$_k$-t$_o$);\linebreak
where f$_o$ is the pulse frequency at time t$_o$ and $\phi$$_o$ is the phase
at time t$_o$). The errors on the Fourier coefficients were corrected for
the non-Poisson noise found in the power spectra \citep{Wilson02}.   In a later 
stage,  a  more precise quadratic-spline phase model was created, which was 
used to  remake the profiles.

Taking consecutive pulse profiles in time during outbursts,
we found a non correlated shape pattern from one outburst to another, and even different patterns either 
going up or down for a given outburst (see Fig.~\ref{consecut_prof}, where  Relative Flux is the 
number of counts per second and per PCU relative to the mean rate). In general, the 
shapes vary from a strong single sinusoidal-like peak to double. However, three or 
even four  weaker peaks are often found 
in addition to the main one. This source showed the same arbitrary conduct 
sorting the profiles by orbital phase (see Fig. ~\ref{same_orb_pha_prof}). Moreover, we 
expected to obtain a base profile for a given  X--ray luminosity.  Nevertheless, we discovered a diversified 
range of shapes  with no clear resemblance (see Fig. ~\ref{same_lumin_prof}). 

Once we realized that no temporal, orbital or luminosity correlations were found in
the original  PCA data sample in the 2-60 keV range, we decided to select only 24 goodXenon 
\rxte PCA observations for the energy dependence study. The chosen data set follows the long-term 
light curve behavior and covers  MJD 51512.8 to  53047.9 (see Fig. ~\ref{Lumin_vs_time}). 
The PCA  background subtracted light curves were obtained 
in four energy ranges: 2.06--5.31 keV, 5.31--7.76 keV, 7.76--13.93 keV, 13.93--20.62 keV. Then, the 
same procedure was followed  to construct the phases and then the profiles, as we described before.
Typically, the duration of a \rxte PCA observation is about 3000 s and the profiles contain about 
8 pulse cycles.
To complete this study, we obtained a list of good events times per science window (ScW)  with IBIS/ISGRI  
in the 20--60 keV energy band.  Table 1  shows a summary of the 
detections and non detections with this instrument, as well as the number of ScWs used 
for making profiles. In Figure ~\ref{Lumin_vs_time} we marked on red (and  an extra circle) 
those 6 points in which we also have an IBIS/ISGRI positive detection. 

\begin{figure*}
  \includegraphics[width=18.5cm,height=8.5cm]{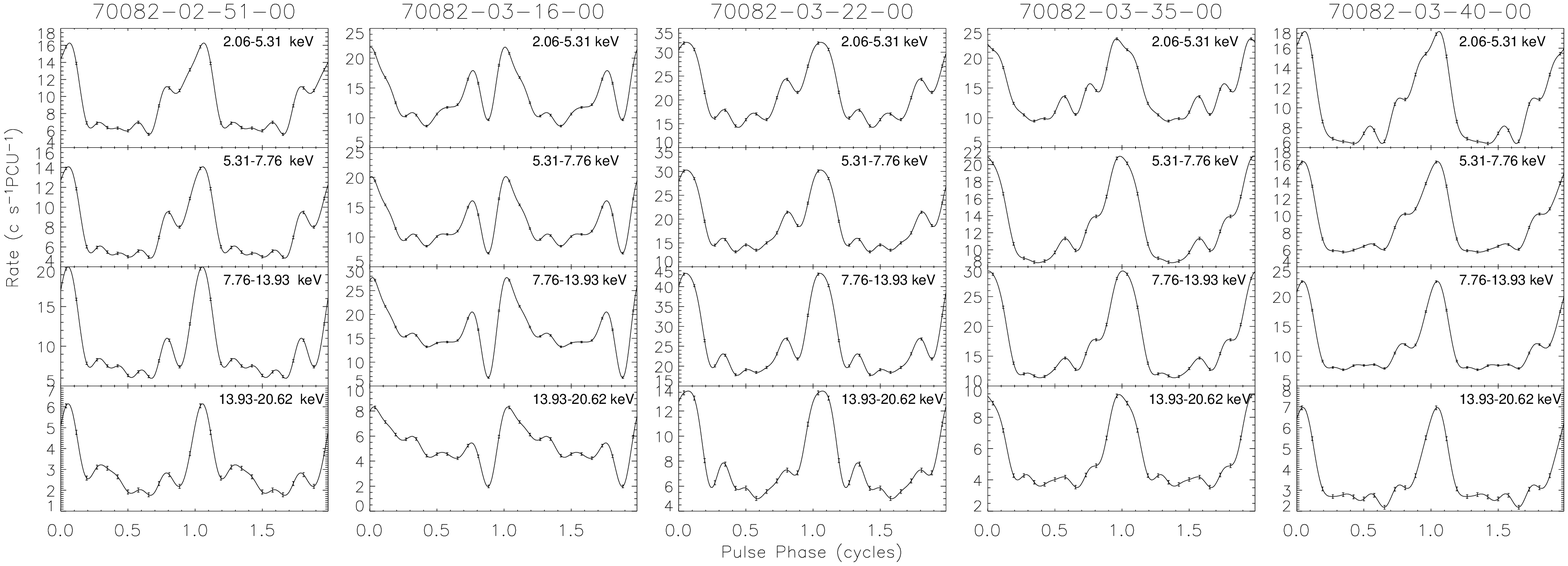}
   \caption{Evolution of SAX J2103.5+4545  PCA pulse profile with intensity in 4 energy bands (from top to
    bottom). They are sorted consecutive in time during the brightest period (MJD 52650--52800) of the source.\label{PCA_prof_1}}
 
\end{figure*}

\begin{figure}
  \includegraphics[width=9.2cm,height=8.2cm]{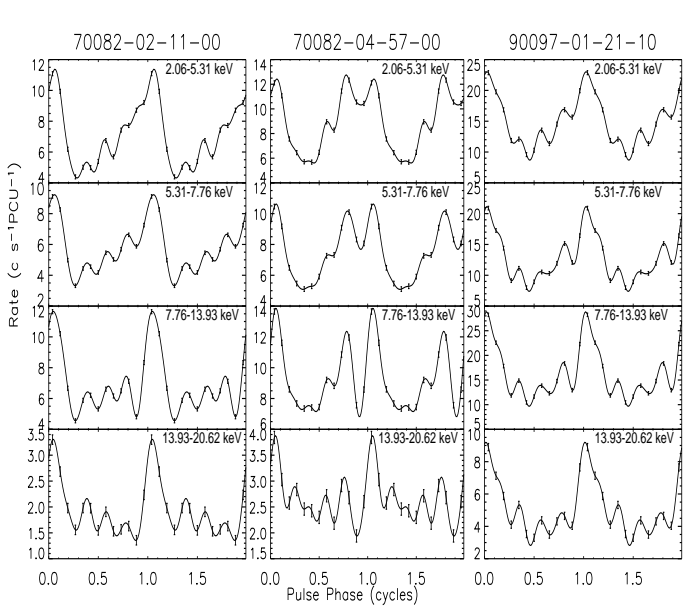}
   \caption{Some examples of  SAX J2103.5+4545  PCA pulse profiles showing a more complex structure, in the same 4 energy bands.\label{PCA_prof_2}}
 
\end{figure}

Figures ~\ref{PCA_prof_1} and ~\ref{PCA_prof_2} show  the evolution of some of the PCA profiles in 
four energy bands (with energy increasing from top to bottom). As we already mentioned,  the 
profiles present  a prominent sinusoidal-like peak  among  other features. These can vary from one 
to three peaks in addition to the main one. The general behavior is:

1. The right side of the main peak (phases 0.0--0.2) remains quite similar as energy increases.

2. The region at phases 0.2--0.5 tends to fill as energy increases. It is not always a peak. 

3. When a peak is present at phases 0.5--0.7, the  larger is the energy  the weaker is the feature.

4. At phases 0.7 to 0.9 a secondary peak or shoulder is present. It gets smaller with increasing 
energy.

5. The dip in the mean peak (phase 0.9) becomes deeper as the flux increases. The maximum depth  is 
found in the 7.76--13.93 keV band, and then decreases again with energy. In particular, observations  
70082-03-16-00 and  70082-04-57-00 show  this notch especially deep and wide, evolving the last one 
to a strong double peak (see Fig. ~\ref{PCA_prof_1} and  ~\ref{PCA_prof_2}).

\begin{figure}[top]
 
   \includegraphics[width=8.6cm,height=6cm]{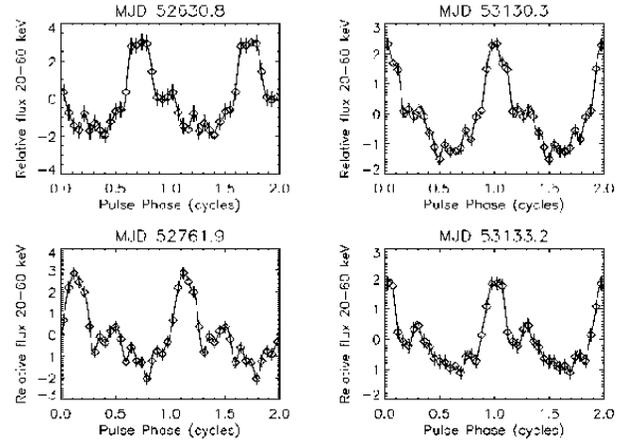}
   \caption{ISGRI (20-60 keV) mean pulse  profiles   of   SAX J2103.5+4545
           generated   by  combining  profiles from individual  \integral  ScWs.
           \label{ISGRI_prof}}
 \end{figure}

This trend is  also seen in the 20--60 keV  IBIS/ISGRI profiles (see Fig. ~\ref{ISGRI_prof}),
where the second and the third peaks are almost missing and the fourth one 
is stronger, but still weaker than the mean peak.

 \begin{figure}
  \includegraphics[width=8.7cm,height=15cm]{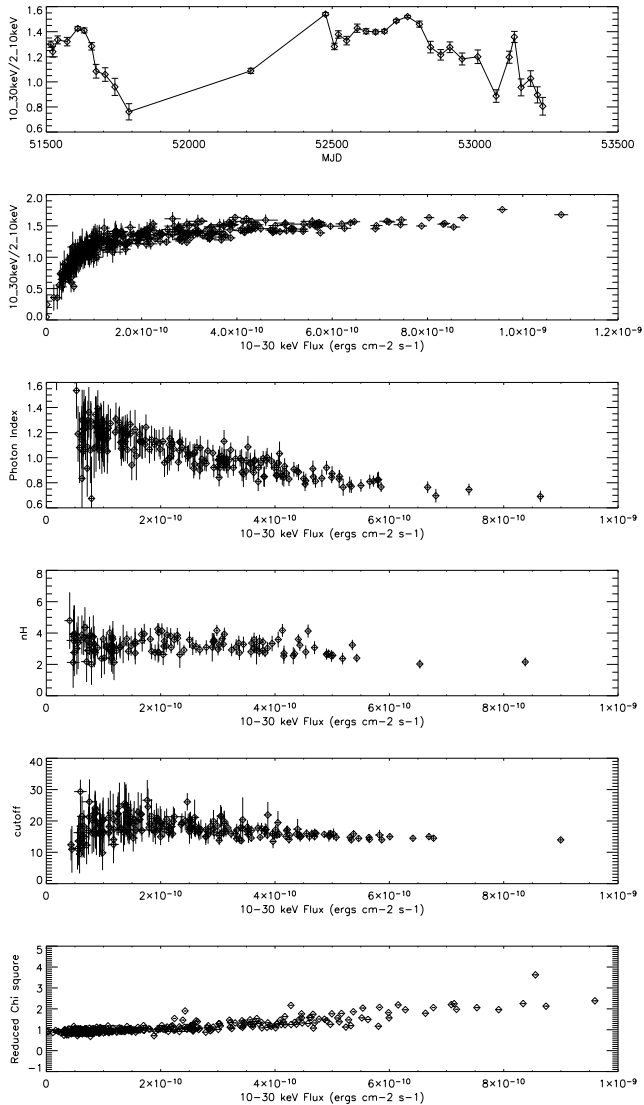}
    \caption{Top. Long-term Hardness ratios  history. From middle to bottom: HR-intensity diagram
     and spectral parameters evolution vs. intensity. \label{HR_study}}
 \end{figure}

 \begin{figure}
  \centering
  \includegraphics[width=8.7cm,height=10cm]{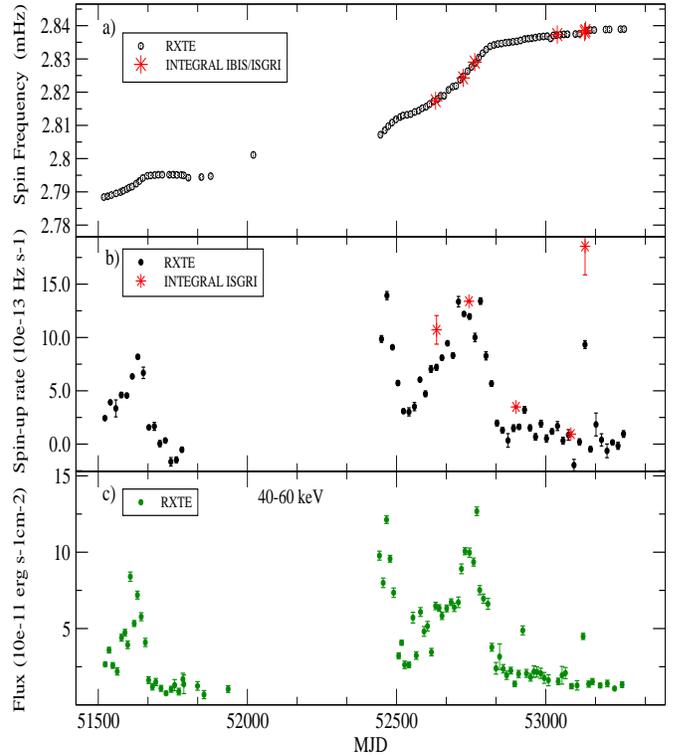} 
   \caption{Top: Long-term frequency history. Middle: Averaged spin-up rates history. 
    Bottom: 40-60 keV \rxte HEXTE  averaged fluxes. A spin-up rate and X--ray flux 
    correlation was found, confirming that an accretion disk during 
    periastron passage is present.\label{flux_fdot_hist}}
 \end{figure}

\subsection{SPECTRAL ANALYSIS}

We have obtained  548 spectra  from 6 years of \rxte PCA and  HEXTE data
(MJD 51512.8 -- MJD 53047.9). For each observation, we have selected PCA 
Standard2 data  and  HEXTE binned mode data from clusters A and B.
Then, fluxes in all the different energy ranges  were  obtained  by  fitting a  
cut-off power law model plus a photoelectric absorption (phabs$\times$(gauss + cutoffpl))
to each  2.7--70 keV spectrum, fixing the iron line value and its width to 6.43 keV and 0.165, respectively 
(Baykal et al. 2002). They were also background subtracted.\linebreak

In fig. ~\ref{flux_fdot_hist} (bottom) we show the long-term flux history of 
SAX J2103.5+4545 using \rxte and \integral data, which follows very nicely the general trend of the 
long-term frequency derivatives history obtained after the timing analysis (see next section).

To study the spectral variability of  SAX J2103.5+4545 we have computed the
long-term hardness ratios (HR) using two energy bands: 2--10 keV, 10--30
keV (defined as $HR= H/S$). Figure ~\ref{HR_study} (top) shows an average of the calculated HR 
vs. time and intensity, using those energy bands. In the HR vs. intensity plot we took out 
few detections corresponding to the faintest state, due to their enormous error bars, which hid the 
general  trend. The correlation between the HR either with time or intensity is clear.
We can see that the source becomes harder when the flux is higher. In the 
same figure we have included some plots showing  the evolution
of the absorption, the Photon Index, the cutoff and the reduced $\chi$$^2$ with intensity. 
We have also rejected the faintest flux values since the uncertainties obtained in the 
parameters were very large, adding only confusion to the study. We found variable behavior 
just in the Photon Index vs. intensity plot. The larger is the flux the lower is the 
the photon index. This anti-correlation was the expected and shows again that 
in the bright state the spectrum of\linebreak  SAX J2103.5+4545 is harder than in the faint state. 
It is clear that the fitting model applied is not the perfect one for
the bright state, as we can see in the reduced $\chi$$^2$ vs. intensity figure. Nevertheless,
it was found to be the best simple model for all the $\sim$500 spectra analyzed.
Baykal et. al (2007) suggested that this type of spectral softening with decreasing flux was found 
to be mainly a consequence of mass accretion rate change and was not necessary to be related 
to a significant accretion geometry change.

\begin{table*}[!] 
\centering  
  \caption{Orbital solution for SAX J2103.5+4545. } 
  \begin{tabular}{llllll} 
   \noalign{\smallskip} 
    \hline \hline \noalign{\smallskip}
                     &    Baykal et al.  &   \.{I}nam et al.   &    Sidoli et al. &    Baykal et al.   & Present work\\
                     &	 (2002)         &     (2004)           &    (2005)        &    (2007)           &         \\
    \hline \noalign{\smallskip}
    T$_{pi/2}$ (MJD)      &  51519.3$\pm$0.2 & 52633.90$\pm$0.05 &      &  52469.336$\pm$0.057    &52545.411$\pm$0.024 \\
     \noalign{\smallskip}
    P$_{orb}$ (days)      &  12.68$\pm$0.25  &       & 12.670$\pm$0.005 & 12.66536$\pm$0.00088    &12.66528$\pm$0.00051  \\
     \noalign{\smallskip}
    a$_x$sin{\it i} (lt-s)&  72$\pm$6        &       &                  & 74.07$\pm$0.86         & 80.81$\pm$0.67\\
     \noalign{\smallskip}
    e                     &  0.4$\pm$0.2     &       &                  & 0.4055$\pm$0.0032      & 0.401$\pm$0.018  \\
     \noalign{\smallskip}
    $\omega$              &  240$\pm$30      &       &                  & 244.3$\pm$6.0           & 241.36$\pm$2.18 \\
    \noalign{\smallskip}
    \hline \hline \noalign{\smallskip} 
   \end{tabular} 
   \label{broad} 
\end{table*} 
\subsection{TIMING ANALYSIS}

\subsubsection{  Phasing}

We selected again the  548 \rxte PCA observations for the long-term 
timing analysis. Then we  obtained either a list of good
events times per science window  (\integral IBIS/ISGRI, 20--60 keV)
or binned light curves (\rxte PCA, 2--60 keV). The times were 
corrected to the barycenter of the solar system 
and for the orbital motion. We constructed pulse profiles in those energy
bands  by fitting the data with a harmonic expansion in pulse phase, as we described
in the Pulse Profile section. Initially we used the same simple phase model and the errors 
on the Fourier coefficients were corrected for the non-Poisson noise.

The profiles were grouped into 12.69 d intervals  with 
frequencies and frequencies derivatives being estimated for each interval
using a grid search of the Y$_n$ statistics \citep{Finger}. The measured
frequencies were then fitted with a piecewise-linear model. Then the profiles
were remade using the integral of the fitted frequency model as a phase model.

For the  \integral IBIS/ISGRI data the approach was different, 
due to the non continuous observational pattern and the sensitivity of the ISGRI 
detector (see Table 1). A template profile was then created from the average profile 
from the set of ISGRI ScWs and from the collected pulse profiles from the \rxte PCA 
data.  To generate phase offsets from the model, we then cross-correlated  the individual profiles 
with the template profiles. The new phases (model + offset) were then fitted with 
a  quadratic-spine phase model, which was used to remake the profiles.
New phase offsets were then estimated.

Figure ~\ref{flux_fdot_hist}  shows  the long-term spin-frequency and spin-up rate history
obtained after repeating the grid search with the new profiles and phases. For IBIS/ISGRI 
the spin rates  were computed by differencing adjacent frequency measurements and dividing by 
the corresponding  time difference. The  PCA spin rates were computed by fitting a quadratic 
function to the phases, which were divided in  29 time intervals. 
We see that the  first  outburst  of\linebreak SAX J2103.5+4545  
started  with a spin-up trend (bright state), 
made a transition to a steady spin rate (faint state), and 
then appeared to just begin a spin-down trend. The following  available  
data started  with a  quick  spin-up trend (bright state), and then a transition 
to a slower spin-up rate (faint state).

Comparing  the fluxes obtained in the non averaged long--term  
light curve (the one composed of around 500 individual flux measurements 
from the spectra; see bottom of fig. ~\ref{Lumin_vs_time}), we realized that while SAX J2103.5+4545 
was in the faint state (MJD 53100), suddenly it increased 
dramatically its  luminosity for a short time period. This was not 
the first episode of such a behavior  but  the strongest one. Therefore, 
the ISGRI frequency derivative outlier cannot be 
ruled out since it seems it is related to a significant event.

\subsubsection{  Orbit fitting}

Errors  in the orbital parameters caused by coupling  between the 
intrinsic spin variations of the pulsar with  orbital effects, can 
scatter the detected pulse frequencies. In order to achieve a better 
characterization of the geometry of this binary system, we have used 
\rxte PCA Standard1 data to obtain a new set of orbital parameters. 
To take into account  the phase noise caused for pulse 
profiles variations we split  the \rxte results into
29 intervals (3 orbits) and fit an  orbit to  the phases in each interval 
(the orbital Period was fixed). The fitting model for each
interval consisted of a orbital model, a  quadratic-phase model.

The final orbital parameters were obtained from a weighted average of the best
fit parameters from just 28 intervals. As we mentioned above,  taking a look at the long-term light curve 
we noticed that around  MJD 53100 took place a huge and narrow spike 
while the source was in the faint state. This surprisingly behavior was also
noticed in the orbital fit number 26, obtaining a highly dispersed fit that was finally
rejected.   The final orbital epoch and period were
obtained by making a linear fit to the 28 best fit orbital epochs. The errors
in these parameters were estimated from the scatter of the best fit parameters.
Table 2 lists the final  orbital parameters  obtained for the combined  
fit, where  T$_{pi/2}$ is the epoch when the mean orbital 
longitude is equal to 90 deg,  P$_{orb}$ is the orbital period, a$_x$sin{\it i}/c is 
the light-travel time for the projected semi-major axis ({\it i} is the 
inclination angle), e is the eccentricity and $\omega$ is the longitude of 
periastron.

\subsubsection{Spin-up Torque-Flux correlations}

\begin{figure}
 \centering
 \includegraphics[width=7.7cm,height=9cm]{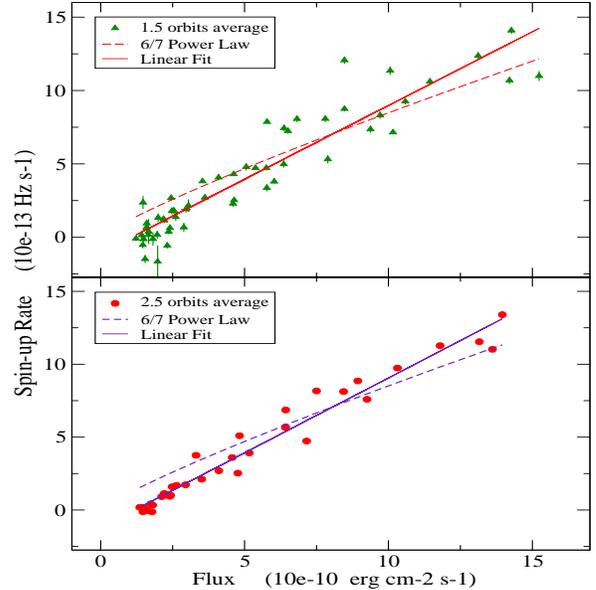}
   \caption{Spin-up rates vs. \rxte  2--100 keV non absorbed averaged Flux every 1.5 and
    2.5 orbits. A linear fit was found the best one.\label{spin_flux_corr}}
\end{figure}

We averaged the non absorbed flux every 1.5 and 2.5 pulsar orbits.  The 1.5 orbits average 
is the smallest time interval that allowed us to fit a quadratic function to the phases when the spin rates
were computed. The second time interval of 2.5 orbits was chosen 
as a good representative of the general trend. Fig. ~\ref{spin_flux_corr}  
shows more clearly the spin-up rate vs. flux correlation,  with 
a linear correlation coefficient of 0.94 and 0.98, respectively. When we fitted both 
data sets with a single power law ({\it solid line}),
we obtained for the best fit  slopes of (1.005--1.027)$\pm$(0.05--0.035)$\times$10$^{-3}$ erg cm$^{-2}$
and an {\it{x}}-intercept (2--100 keV non absorbed flux at zero spin-up) of 
(-1.07--(-1.21))$\pm$(0.32--0.22)$\times$10$^{-10}$ ergs cm$^{-2}$ s$^{-1}$. Both  results are
coherent. They are identical within errors, however it is interesting to note that data
averaging over 1.5 orbits scatters more than averaging over 2.5 orbits. It appears that the smaller is the 
average the bigger is the scattering. This might suggest that, in addition to 
the disk accretion mechanism, there seems to exist another mechanism  producing erratic variations.

In the same figure, a dashed line denotes our best-fit power 
law with an index fixed at 6/7, which represents the relationship
between spin-up and flux predicted by simple accretion theory. We can see that this fit does 
not match totally the observed behavior. It suggests that a more sophisticated model, e.g. Ghosh
and Lamb, as in Baykal et al. (2007), seems to be needed at low-flux end,  because the pulsar is 
beginning  to spin-down. At the high-energy flux end, beaming may be boosting the observed flux.

\section{DISCUSSION}

\subsection{NATURE OF THE SOURCE}

Some X--ray pulsars, like EXO 2030+375 (Parmar et al., 1989), 
show a clear pulse shape dependence on source luminosity, most likely controlled by accretion rate.
Differences in X--ray pulse profiles are considered to be due to the differences in the 
geometrical configuration with respect to the rotational axis of the neutron star, the axis of
magnetic dipole moment, and the observer's line of sight (Nagase 1989). The structure 
of the accretion column determines the basic profile of the pulse pattern. However, 
simple geometric models with two pencil beams coming from the magnetic poles
or two bright spots on the neutron star surface, cannot explain the variability 
observed in some other systems, like Vela X--1.\linebreak Previous
studies by Kreykenbohm et al. (2002) confirmed for that system a very complex 
pulse profile at low energies (below 6 keV), showing a five peak structure. Above 
10 keV the pulse profiles evolved into a simple double peak. They pointed out that
a good description of the processes responsible for the complex shape at low energies 
is still missing.

In the present work, the  2--60 keV  SAX J2103.5+4545 pulse profiles obtained  with \rxte PCA
were found to be peculiarly complex and variable either with luminosity, time or orbital phase.
For the first time the pulse  shapes seem to vary randomly 
from not only single sinusoidal-like peak but  profiles with two, tree and even occasionally four peaks,  
with no evident  interrelationship. For a given energy range, temporal variability of the pulse profile 
of SAX J2103.5+4545 was explained by Sidoli et al. (2005) likely  due 
to time-dependent emission pattern, or to changes in the opacity of the 
magnetized plasma where the radiation propagates.

Previous works by Baykal et al. (2002),  Falanga et al. and  Sidoli et al. (2005) obtained a 
single peak pulse profile in soft X--ray bands, using \rxte and \integral data.  
However, \.Inam et al. (2004) found a double peak profile using  {\it XMM-Newton}. Although 
this seem not  to be in agreement, could be compatible due to the variability seen in our 
study for SAX J2103.5+4545. 
 
In the hard X--ray range, we have obtained much more simple 20--40 keV \integral 
ISGRI profiles. In particular, double-like peak profiles  
in agreement with previous  results by  Falanga et al. and  Sidoli et al. (2005).
Filippova et al. (2004) published a 20--100 keV  single peak pulse profile using 
PV phase \integral data. This result might be  in  agreement with our PV phase 
results, taking into account their uncertainties.

In addition, our study allowed us to confirm an energy dependence pattern for SAX J2103.5+4545. 
Unfortunately, as in Vela X--1, we did not find out a straightforward explanation for 
the low energy pulse profiles. \.Inam et al. (2004) found  practically  no 
variation in the energy dependence of the 0.9--11 keV
pulse profiles, which is not in agreement with our results. Falanga et al.   (2005) 
obtained an energy dependent secondary peak around phase 0.2, which became
more evident at energies above 20 keV. This peak might be compatible with the trend of our peak
at phase 0.7. They also claimed that at phase 0.6 hick-ups were present, in agreement with
those observed by  \.Inam et al. (2004) using {\it XMM-Newton} data. These hick-ups 
were only visible at high energies (40--80 keV). Again, these features might be compatible 
with our dip at phase 0.9.

It is generally accepted that in order to explain the 
energy-dependent changes of the pulse profile, the anisotropic  
radiation transfer must be taken into account (Nagase 1989). Previous studies by Krauss et al. (2003) found that 
energy-dependent peaks in medium luminosity binary X--ray pulsars, 
are mainly due to energy-dependent relative importance of the halo (which forms around the accretion 
funnel where the neutron star surface is irradiated) and the column contributions to 
the observed flux \cite{Sid1}. For SAX J2103.5+4545, Falanga et al. (2005) found that 
changes in the morphology of the pulse profile as a function of energy were  consistent 
with variations in the spectral components  visible in their  pulse phase resolved spectra 
analysis. Their study  also showed that differences in the double-peak  can be modeled by 
a different scattering fraction between the radiation from the two magnetic poles.

First  studies of SAX J2103.5+4545 by Baykal et al. (2002) showed a transition from spin-up to 
spin-down  in the \rxte observations from an outburst in  November 1999, suggesting 
the presence of an accretion disk.  The detection of a quasi-periodic oscillation 
at 22.7 s discovered by  \.{I}nam  et al.  in 2004,  provided further evidence
for this existence.  The spin-up rate and X-ray flux correlation  
is also observed in the present work, confirming that an accretion disk is present during 
periastron passage. Previous works  by \.{I}nam et al. (2004) and Baykal et al. (2002, 2007) are
in good agreement with our results.  In general, either the pulse  period  
estimations  from Sidoli et al. (2005) or those ones from  Blay et al. (2006) are 
not in very good agreement with our results.

It is frequently assumed that in normal outbursts no disk forms around the neutron star and
accretion is directly from the disk-like outflow from the Be star, so significant spin-up 
is not expected because direct wind accretion is not believed to be very efficient at 
transferring angular momentum (Wilson et al. 2002, and references therein). If enough angular momentum
is present in the accreted material, an accretion disk will form. However, evidence
for spin-up during normal outbursts has been observed in \linebreak GS 0834-430 (Wilson et al. 1997),
2S 1417-624 (Finger, Wilson $\&$ Chakrabarty 1996), 2S 1845-024 (Finger et al. 1999),
and in EXO 2030+375 (Stollberg et al. 1999; Wilson et al. 2002, 2005).

Simple accretion theory assumes that the material from the companion star is flowing onto a 
rotating neutron star with a strong magnetic field. This field determines the motion of 
material in a region of space called the magnetosphere. The size of this region is denoted 
by the magnetospheric radius r$_{m}$, given by (Pringle $\&$ Rees 1972; Lamb, Pethick $\&$ Pines 1973)
 
\begin{equation}\label{rm}
r_{m} \simeq \it {k}\it {(GM)}^{-1/7}\mu^{4/7}{\dot{\it {M}}}^{-2/7}
\end{equation} 

where {\it {G}} is the gravitational constant, {\it {k}} is a constant factor of order 1, {\it {M}} is the mass 
of the neutron star, and {\it {\.{M}}} is the mass accretion rate: {\it {\.{M}}} 
= 4$\pi${\it {d}}$^2${\it {FR/GM}} (where {\it {R}} is the radius of the neutron 
star and {\it {d}} is the distance to the pulsar).
Equation (1) with {\it {k}} $\simeq$ 0.91 gives the Alfv\'en radius for spherical accretion and with 
{\it {k}} $\simeq$ 0.47{\it {n($\omega$$_{s}$)}} gives the magnetospheric radius derived by Ghosh $\&$
Lamb (1979).

The torque applied by accretion of matter onto a neutron star, assuming torques due to matter 
leaving the system are negligible, is given by (Lamb, Pethick $\&$ Pines 1973)

 \begin{equation}\label{dl_dt}
\frac{\it {d}} {\it {dt}}
 {(2 \pi\it {I}\nu)} = \dot{\it {M}}\it {\ell}         
\end{equation}   

where {\it {I}} is the moment of inertia of the neutron star and {\it {$\ell$}} is the specific angular 
momentum of the material. If {\it {I}} is assumed constant, then  {\it {$\ell$}} is given by
\begin{equation}\label{l}   
\it {\ell} = 2 \pi {\it {I}} \dot{\nu} \dot{\it {M}}^{-1}
\end{equation}

where $\dot\nu$ is the spin-up rate. To estimate {\it {$\ell$}} for\linebreak SAX J2103.5+4545, we assumed 
typical pulsar parameters, {\it {M}} = 1.4 {\it {M}}$_{\odot}$, {\it {R}} = 10 km, 
{\it {I}} = 10$^{45}$ g cm$^{2}$, typical values of $\dot\nu$ $\simeq$ 7 $\times$10$^{-13}$ Hz s$^{-1}$,
{\it {F}} $\simeq$ 7 $\times$10$^{-10}$ erg cm$^{-2}$ s$^{-1}$ (see Fig. ~\ref{spin_flux_corr}).
The distance of the source from optical observations is 6.5 $\pm$0.9 kpc (Reig et al., 2004, 2005). 
Several authors have discussed a possible value for the magnetic field. Recently, Baykal et al.  (2007) 
obtained a value of 16.5$\times$10$^{12}$ Gauss. But this implied a distance of 4.5 kpc. In order to explain 
the spin-up rate observed, and  assuming  that the distance of 6.5 kpc is correct, Sidoli et al. (2005) 
obtained  a magnetic field of $\sim$ 1.6$\times$10$^{12}$ Gauss. No 
cyclotron lines have been observed from this source. One explanation could be a high magnetic field 
as proposed by Baykal et al. (2007) which would imply a fundamental line at $\sim$200 keV.

An accretion disk will form if the specific angular momentum of the material accreted from the Be star's disk is
comparable to the Keplerian specific angular momentum at the magnetospheric radius, i.e.
\begin{equation}\label{lm} 
 {\it {\ell}} \simeq   {\it {\ell_{m}}} = (\it {GMr_m})^{1/2}
\end{equation}

For SAX J2103.5+4545, we have summarized in Tab. 3 the different values obtained for{\it {$\ell$}},
using the two  magnetic field measurements discussed above,  assuming two 
distance values of 6.5 kpc and 4.5 kpc, and for\linebreak  {\it {k}} = 0.91 and  0.47.
All the cases suggest that a disk is likely present. Only those cases which lower 
 {\it {$\ell$}}/{\it {$\ell$$_m$}} might indicate that periods of wind accretion and 
disk accretion could also take place in this system. 

In contrast, for the wind-fed system Vela X--1 where a disk is not expected to be present,
$\dot\nu$ $\simeq$ 6 $\times$10$^{-14}$ Hz s$^{-1}$,\linebreak
{\it L } $\simeq$ 2$\times$10$^{38}$ ergs s$^{-1}$, and $\mu$ $\simeq$ 2.1 $\times$10$^{30}$ G cm$^3$
leading to {\it {$\ell$}} $\simeq$ 0.002{\it {$\ell$$_{m}$}}. Furthermore, three-dimensional 
simulations of wind accretion show that the average specific angular
momentum accreted via this mechanism is always smaller than the keplerian value 
(Wilson et al. 2003, and references therein).


\begin{table} 
  \caption{Table 3. Comparison of the specific angular momentum of the accreted material
                    to the Keplerian specific angular momentum at the magnetospheric radius. An 
                    estimation of the flux for the onset of centrifugal inhibition of accretion is also shown.  } 
  \begin{tabular}{p{1.3cm} l l} 
  \noalign{\smallskip} 
    \hline \hline\noalign{\smallskip}
                    & {\it B}= 16.5$\times$10$^{12}$  G                      &  {\it B}= 1.65$\times$10$^{12}$ G  \\
                    &     (4.5 -- 6.5) kpc                                   &    (4.5 -- 6.5) kpc                          \\
   \hline \noalign{\smallskip}
    {\it {k}}=0.91  &  {\it {$\ell$}}= (0.61 --  0.33){\it {$\ell$$_{m}$}}   &   {\it {$\ell$}}= (1.19 --  0.63){\it {$\ell$$_{m}$}} \\
                    &  F$_{min}$$^*$=  (20.4 -- 9.8)                         &   F$_{min}$$^*$=   (0.2 -- 0.097)\\
    {\it {k}}=0.47  & {\it {$\ell$}}= (0.85 --  0.45) {\it {$\ell$$_m$}}     &   {\it {$\ell$}}= (1.65--  0.88) {\it {$\ell$$_{m}$}} \\
                    &  F$_{min}$$^*$= (2.02 --  0.97)                        &   F$_{min}$$^*$=   (0.02 -- 0.0097)\\
    \hline \hline \noalign{\smallskip} 
       & $^*$ $\times$10$^{-12}$  erg cm$^{-2}$s$^{-1}$ & \\

   \end{tabular} 
   \label{angular} 
\end{table} 


To determine whether or not  centrifugal inhibition of accretion is operating, we estimated 
the flux at the onset of this effect by equating the magnetospheric radius to the corotation radius. 
The magnetospheric radius is given by equation ~\ref{rm}, and the corotation radius is given by 

\begin{equation}\label{r_co}
r_{co} = \it {(GM)}^{1/3} (2\pi\nu)^{-2/3}
\end{equation} 

where $\nu$ is the spin frequency of the pulsar. Setting\linebreak r$_m$ = r$_{co}$ gives the threshold flux for 
the onset of centrifugal inhibition of accretion, i.e.,

\begin{equation}\label{f_min}
\begin{split}
  {\it F_{min}} \simeq 2\times10^{-12} erg cm^{-2} s^{-1} \\
\times {\it k}^{7/2} {\it \mu_{30}}^{2} {\it M_{1.4}}^{-2/3} {\it R_6}^{-1}
{\it P_{354.9s}}^{-7/3} {\it d_{kpc}}^{-2} 
\end{split}
\end{equation}

where $\mu$$_{30}$, {\it M$_{1.4}$}, {\it R$_6$}, and {\it P$_{354.9}$} are the pulsar's magnetic moment 
in units of 10$^{30}$ G cm$^{3}$, mass in units of 1.4{\it {M}}$_{\odot}$, radius in units of 10$^{6}$ cm, 
and spin period in units of 354.9 s, respectively.  Tab. 3 shows our measured  F$_{min}$ using
equation ~\ref{f_min}. Our observed  upper limit fluxes are  in the range  
$\simeq$  (5--0.9)$\times$10$^{-12}$  erg cm$^{-2}$s$^{-1}$,  consistent with SAX J2103.5+4545 entering 
the centrifugal inhibition of accretion regime only for {\it B}= 16.5$\times$10$^{12}$ G (see Tab. 3).

\subsection{TRANSIENT  VS. PERSISTENT}

Up to now, SAX J2103.5+4545  has been classified as a transient Be/X--ray 
binary system.  However, this source is a very interesting case. A study by 
Reig et al. (2005) of the correlated X/optical data of this source during quiescence, 
showed that X--ray emission  must come from stellar wind, since the source had completely lost 
its circumstellar disk. He also pointed out that  SAX J2103.5+4545 occupies the 
region of the wind-fed supergiant binaries in the\linebreak P$_{spin}$-P$_{orb}$ diagram. Thus, in 
principle,  accretion  from the stellar wind of the BO companion might be at the 
origin of the observed luminosity in that state. And maybe, to be located in the 
wind-fed supergiant region might be also the origin of the complex SAX J2103.5+4545 
pulse profile behavior.


 \begin{figure}
 \centering
 \includegraphics[width=8.4cm,height=4cm]{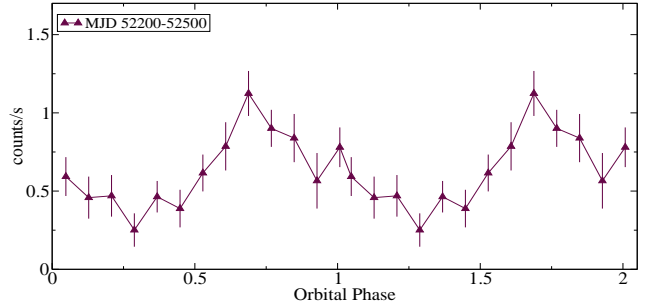}
   \caption{\rxte ASM orbital light curve of SAX J2103.5+4545 from  MJD 52200--52500.\label{asm_folded_lc}}
\end{figure}


In the present work we have not obtained observations where 
SAX J2103.5+4545 was not detected, hence it is unclear how to determine  outbursting vs. quiescence.
Defining a threshold for the observed flux at 5$\times$10$^{-12}$  erg cm$^{-2}$s$^{-1}$, we 
might see 5 outbursts from Fig. ~\ref{Lumin_vs_time} (bottom).    Ignoring the fact that the 
luminosity never exceeds 10$^{37}$  erg cm$^{-2}$s$^{-1}$, the first three might be qualify as 
type II outbursts: a) they last for multiple orbits,  b) there is only moderate orbital modulation 
(see Fig. ~\ref{asm_folded_lc}),  c) there is a flux-spin up  correlation. The last two outbursts
are short. In the ``quiescence'' state where the luminosity is
 $\sim$10$^{35}$ erg cm$^{-2}$s$^{-1}$,  we see a gradual flux decline after the third ``type II''
outburst, yet the frequency continues to increase. This is again  not typical, but it has been seen
in wind-fed systems (Bildsten et al, 1997). Overall,  the behavior 
of the source does not fit well with the standard picture of Be/X-ray transients. 
We also believe that this peculiar Be/X--ray system with the narrowest orbit, may be 
classified as a Persistent (but highly variable) source.
\linebreak
\linebreak

{\it Acknowledgments}. We thank Peter Kretschmar for very useful comments and his support. We also 
appreciate to NASA Marshall Space Flight Center, the National Space Science Technology Center 
and the  Universities Space Research Association for the opportunity  to develop the present work. 
This research is supported by the Spanish Ministerio de Educaci\'{o}n y Ciencia 
 through grant-no ESP2002-04124-C03-02.

\end{document}